# The microscopic nature of localization in the quantum Hall effect

S. Ilani[1*], J. Martin[1], E. Teitelbaum[1], J. H. Smet[2], D. Mahalu[1], V. Umansky[1] & A. Yacoby[1]

[1]*Department of Condensed Matter Physics, Weizmann Institute of Science, Rehovot 76100, Israel*

[2]*Max-Planck-Institut für Festkörperforschung, D-70569 Stuttgart, Germany*

[*]*Present address: Laboratory of Atomic and Solid State Physics, Cornell University, Ithaca, NY 14853, USA*

**The quantum Hall effect arises from the interplay between localized and extended states that form when electrons, confined to two dimensions, are subject to a perpendicular magnetic field[1]. The effect involves exact quantization of all the electronic transport properties due to particle localization. In the conventional theory of the quantum Hall effect, strong-field localization is associated with a single-particle drift motion of electrons along contours of constant disorder potential[2]. Transport experiments that probe the extended states in the transition regions between quantum Hall phases have been used to test both the theory and its implications for quantum Hall phase transitions. Although several experiments[3–9] on highly disordered samples have affirmed the validity of the single-particle picture, other experiments[10–12] and some recent theories[13–15] have found deviations from the predicted universal behaviour. Here we use a scanning single-electron transistor to probe the individual localized states, which we find to be strikingly different from the predictions of single-particle theory. The states are mainly determined by Coulomb interactions, and appear only when quantization of kinetic energy limits the screening ability of electrons. We conclude that the quantum Hall effect has a greater diversity of regimes and phase transitions than predicted by the single-particle framework. Our experiments suggest a unified picture of localization in which the single-particle model is valid only in the limit of strong disorder.**

Unlike extended electrons, whose charge is spread over the entire volume of the system, a localized electron is confined to a small region in space, determined by its localization length. Upon population, a single electronic charge enters this confined region and results in a discrete jump in the local chemical potential $\mu$ and a spike in its derivative with respect to the back-gate voltage $d\mu/dV_{BG}$. Hence, the local derivative

d$\mu$/d$V_{BG}$, which is inversely proportional to the local electronic compressibility[16], provides direct information about the localized states within quantum Hall (QH) phases. To measure this derivative, we use a single-electron transistor (SET) electrometer[16–20]. Figure 1a shows a coarse measurement of d$\mu$/d$V_{BG}$ as function of magnetic field (*B*) and density (*n*), which is tuned by the back-gate voltage. Incompressible behaviour appears along integer and fractional values of $\nu=hn/eB$ (bright strips) corresponding to integer and fractional QH phases (here *h* is Planck's constant, and *e* is the charge on an electron). A detailed measurement of d$\mu$/d$V_{BG}$ within the QH phases (Fig. 1b) reveals a rich fine structure of parallel (black) lines. Each line is the evolution in the *B*–*n* plane of a single spike in d$\mu$/d$V_{BG}$ and corresponds to charging of an individual localized state.

Figure 1b shows two distinct groups of localized states that surround the $\nu=1$ and $\nu=2$ QH phases. Similar groups appear around every well-developed integer and fractional QH phase. In addition, a group of horizontal lines that are associated with the insulating phase ($\nu=0$) is observed at low densities (Fig. 1a). Further examination of these groups reveals three interesting properties. First, the number of states within each group is independent of *B*. This is surprising, because, for example, a twofold increase in *B* results in a twofold increase in the Landau level degeneracy, but hardly changes the number of localized states we observe (Fig. 1c). Second, the groups of localized states form strips of constant width, $\Delta n$, in the *n*–*B* plane, and $\Delta n$ is the same for different QH phases (Fig. 1a). In fact, when the longitudinal resistivity ($\rho_{xx}$) is measured in the same sample, QH behaviour (zero $\rho_{xx}$) appears also along strips with similar $\Delta n$ (Fig. 1d). The quantitative correspondence between the transport and local compressibility implies that we probe most or all of the localized states that are responsible for the QH phenomena in transport. Third, localized states that belong to the same group evolve in the *n*–*B* plane with exactly the same slope: a slope equal to the slope of their underlying QH phase. Thus, the slope of the charging lines is quantized to d$n$/d$B=e\nu/h$ ($\nu$ integer or fractional). The slope, therefore, identifies each localized state with the QH phase it belongs to, and it remains distinctively quantized even in regions where two groups of states overlap (Fig. 1e). Similar phenomenology is observed in all the samples we have studies (four different samples), which cover a large range of peak mobilities ((0.5–4)×10$^6$ cm$^2$ V$^{-1}$ s$^{-1}$).

The measured pattern of localized states is summarized schematically in Fig. 2a. States are bunched in groups that correspond to the various phases of the system. Each group contains a constant number of lines with identical slope. We will now show that these observations are inconsistent with the single-particle description of localized states. Imagine a model disorder potential in the shape of a Mexican hat (Fig. 2a inset). In the single-particle picture, the localized states are constant-energy orbits and enclose integer number of flux quanta. For our model potential, these orbits form concentric circles (blue lines). When *B* increases, all orbits shrink to maintain constant flux through their area (magenta lines). Thus, at higher *B* there are more states per unit area. This is a generic property of the single-particle picture, which results from the enhanced Landau level degeneracy as *B* increases. A microscopic compressibility measurement at any arbitrary location would therefore reveal an increasing number of localized states as the degeneracy of the Landau levels increases, which is in sharp contrast to the constant number of localized states observed. Furthermore, if we evaluate in our simple model the energy of an individual state as function of *B* (for example, the state shown green) we notice that it is non-monotonic—first going down to the bottom of the potential landscape and subsequently climbing up the central hill. Correspondingly, its charging line in the *n*–*B* plane (green line in Fig. 2a) first descends and then rises with a slope that asymptotically approaches a non-quantized value. This non-universal behaviour of the charging lines is another generic property of single-particle states and is in sharp contrast with our observations.

Single-particle physics fails to describe our observations because it ignores the interactions between electrons and hence screening. Electrons in this single-particle picture populate states in a fixed disorder potential. However, Coulomb repulsions produce a potential landscape that varies with the filling factor in an attempt to minimize the total electrostatic energy[21]. Such self-consistent potential landscape is a result of spatial density fluctuations, formed by electrons to oppose the bare disorder potential, $V_{bare}(r)$, where *r* is position in real space. For any $V_{bare}(r)$ a specific density profile exists, $n_{scrn}(r)$, which screens it completely and thereby minimizes the electrostatic energy. This disorder-induced density profile, which follows from the Poisson equation $\nabla^2 V_{bare}(x,y,z) = -4\pi e n_{scrn}(x,y)\delta(z)$, has a typical amplitude $\Delta n_{disorder}$, and is uniquely defined up to an arbitrary constant, which we shall take to be zero. We will show that the evolution of the electronic states is governed mainly by the ability or inability of the electron gas to form this screening density profile.

An intuitive picture of localization induced by interaction is obtained by tracing the self-consistent potential as a function of *B* and *n*. Figure 2b depicts the calculated density profiles and the corresponding self-consistent potentials obtained rigorously within the Thomas–Fermi approximation[22] for an almost-empty (I), half-full (II), and nearly-full (III) lowest Landau level with typical disorder. For a given *B* the density in the Landau level cannot exceed one electron per flux quanta, $n_{max}=B/\phi_0$, and is therefore constrained by $0 \leq n \leq n_{max}$. At the centre of the Landau level this constraint is insignificant because the variations in density required for screening ($\Delta n_{disorder}$) are smaller than $n_{max}$. The density follows the profile required for perfect screening, $n(r) = n_{scrn}(r) + \langle n \rangle$, where $\langle n \rangle$ is the average density, thus leading to a flat electrostatic potential (II in Fig. 2b). Each electron added to the system experiences this flat potential and thus, within this approximation, is completely delocalized. Hence, as the density increases, $n(r)$ grows uniformly while maintaining its spatial dependence to ensure perfect screening. Electrons fail to flatten the bare disorder potential when the desired $n(r)$ for perfect screening exceeds $n_{max}$. This commences near the peaks of $n(r)$, leading to the formation of local incompressible regions with a full Landau level ($n(r)=n_{max}$) where the unscreened bare potential sticks out (III in Fig. 2b). These regions coexist with compressible regions where the Landau level is not yet filled ($n(r)<n_{max}$), having a screened potential. Once a compressible region becomes surrounded by an incompressible barrier it behaves as a quantum dot whose charging is governed by Coulomb-blockade physics, that is, electrons are localized to this region and enter it one by one.

The charging spectra of such interaction-induced localized states are very different from those emanating from the single-particle picture, and account for all the properties of localized states we observed. For simplicity we describe in more detail the states around $\nu=1$. Along the $\nu=1$ line, where the first Landau level is exactly full, the bare potential is unscreened and the two-dimensional electron system (2DES) is incompressible everywhere. Reducing the average density introduces holes in the Landau level that accumulate at compressible pockets $(\delta n(r) \equiv n(r) - n_{max} < 0)$ surrounded by an incompressible QH phase $(\delta n(r) = 0)$. These pockets behave as quantum antidots (or dots for $\delta n(r) > 0$). For a given antidot, holes are added at discrete values of local density deviation, $\delta n_i$, and a specific number of holes can be introduced before the confining incompressible barriers disappear. The values $\delta n_i$ are determined solely by electrostatics and are independent of $B$. The only effect of $B$ is to increase $n_{max}$ thereby converting each charging point into a line in the $n$–$B$ plane. This results in a constant number of charging lines running parallel to the $\nu=1$ line, in accordance with our measurements. We note that the formation of localized states is determined solely by the density deviation from a completely full Landau level, $\delta n$, and not by the total density, $n$. Hence, despite the increase of Landau level degeneracy ($n_{max}$) with $B$, the number of localized states remains constant.

Dots also emerge when electrons start filling an empty Landau level (I in Fig. 2b). Here compressible pockets $(\delta n(r) \equiv n(r) - 0 > 0)$ are enclosed by incompressible empty regions $\left(\delta n(r) = n(r) = 0\right)$, resulting in charging lines that are parallel to $\nu$=0. Hence, it is the incompressible phase surrounding a dot that determines the quantized slopes of its localized states. In the experiment, we clearly resolve the boundaries where localized states cease to exist. They appear when the compressible dots form a connected network, lose their quantum confinement, and therefore lose their discrete charging spectra. This percolation transition of a compressible medium in an incompressible background[21,23,24] occurs when the density of compressible fluid equals $\delta n = \Delta n_{disorder}/2$. Within each Landau level two such percolation transitions take place, one for electrons, when $n = \Delta n_{disorder}/2$ and one for holes, when $n = n_{max} - \Delta n_{disorder}/2$. This explains why localized states and QH transport phenomena appear only in strips around quantized values of $\nu$ and associates the width of the strips with $\Delta n_{disorder}$. The observation that different QH phases display the same width in density is thus a natural consequence of the dot model. The number of localized states within the strips is independent of $B$. At higher $B$, new electrons add up to each Landau level owing to its enhanced degeneracy, but these electrons populate non-localized states within the percolating compressible regions, keeping the number of localized states fixed.

Further corroborating evidence for the dot model comes from spatial imaging of the localized states. A SET is mounted on a tip of a scanning microscope (Fig. 3a) and $d\mu/dV_{BG}$ is measured as function of position ($x$) and density at temperature $T$=0.3 K. In the $n$–$x$ plane each localized state appears as a charging segment, whose length in $x$ represent the spatial extent of the localized state convoluted with the tip response function (Fig. 3b). A spatial scan for nearly full lowest Landau level is plotted in Fig. 3c. Charging segments can be seen below and above the $\nu$=1 line. Localized states are not distributed randomly, but rather observed at specific locations with equally spaced charging spectra. These spectra represent generic behaviour observed at all positions across the sample and in different samples, covering a large range of mobilities. They directly manifest the presence of quantum dots—each dot gives rise to a train of Coulomb-blockade peaks at a certain position with level separation given by its charging energy, determined primarily by its area.

In the dot model, the spectra of localized states are determined only by the bare disorder potential and the presence of an energy gap. To verify this prediction we compare in Fig. 3c and d the spectra measured at the same position around gaps of totally different nature: the spin gap at $\nu$=1 and the orbital gap at $\nu$=2. The similarity is evident—both dot and antidot appear in the scans at exactly the same locations, accommodate the same number of electrons and possess the same level spacing.

A different regime is entered when $n_{max}$ becomes smaller than the size of density fluctuations in the 2DES ($\Delta n_{disorder}$). In any given sample, this occurs for low enough $B$. In this highly disordered regime, the separation between groups of localized states ($n_{max}$) drops below their width ($\Delta n_{disorder}$) and charging lines from different groups start to overlap. Experimentally, rather than merging or bending, lines from different groups maintain their original slope and coexist (Fig. 1e). This coexistence implies that phase mixing might exist on microscopic scales. To capture the microscopic picture in this regime we follow in Fig. 4a the $\nu=1$ antidot of Fig. 3 down to low magnetic fields. As $B$ decreases its spectrum shifts down with the top of the first Landau level (top density in all figures) while retaining its internal structure. To the left, another spectrum of a $\nu=0$ dot is visible, which remains intact upon changes in $B$. The fact that the internal structure of both the $\nu=0$ dot and the $\nu=1$ antidot are independent of $B$ demonstrates once more that these spectra are mainly determined by charging energy and not single-particle localization. At the lowest values of $B$, the dot and antidot coexist at the same electron density and their localized states are less than 1 μm apart. Figure 4b shows a self-consistent calculation of $n(r)$ in this highly disordered regime, to clarify that this is indeed plausible. The small degeneracy of the Landau level, $n_{max}$, chops off the density profile ($0 \leq n \leq n_{max}$). The sharp jumps from empty to completely full Landau level result in the coexistence of $\nu=0$ and $\nu=1$ regions, and compressible pockets in these regions produce $\nu=0$ dots that neighbour $\nu=1$ antidots. A new type of compressible strips appear at the boundaries between the $\nu=0$ and $\nu=1$ regions[13–15]. These strips follow the contours of the bare disorder potential (Fig. 4b), reminiscent of single-particle states. In the strong disorder limit, only these compressible strips survive and generate single-particle-like behaviour.

We summarize our results in a generalized phase diagram of localization in the QH regime (Fig. 4c). The disorder in any given sample is characterized by $\Delta n_{disorder}$ irrespective of the density. Depending on the interaction strength relative to the disorder, localized states acquire a different nature. When disorder is strong or when $B$ is small ($\Delta n_{disorder} \gg n_{max}$), localized states are narrow compressible strips, which obey the single-particle phenomenology. Consequently, the familiar QH phase transitions exist at half-filling of the Landau levels. This explains why single-particle scaling of QH transitions is observed only in highly-disordered samples[3–9]. At higher magnetic fields, localization is dominated by interactions and localized states form within dots or antidots. In this regime, new percolation phase boundaries emerge (red lines in Fig. 4c), limiting the QH phenomena to strips of width $\Delta n_{disorder}$ centred around the QH filling factors.

1. Prange, R. E. & Girvin, S. M. *The Quantum Hall Effect* (Springer, New York, 1990).

2. Huckestein, B. Scaling theory of the integer quantum Hall-effect. *Rev. Mod. Phys.* **67,** 357–396 (1995).

3. Wei, H., Tsui, D., Paalanen, M. & Pruisken, A. Experiments on delocalization and universality in the integral quantum Hall effect. *Phys. Rev. Lett.* **61,** 1294–1296 (1988).

4. Wei, H. P., Lin, S. Y., Tsui, D. C. & Pruisken, A. M. M. Effect of long-range potential fluctuations on scaling in the integer quantum Hall-effect. *Phys. Rev. B* **45,** 3926–3928 (1992).

5. Engel, L., Wei, H. P., Tsui, D. C. & Shayegan, M. Critical exponent in the fractional quantum Hall-effect. *Surf. Sci.* **229,** 13–15 (1990).

6. Koch, S., Haug, R., von-Klitzing, K. & Ploog, K. Size-dependent analysis of the metal-insulator transition in the integral quantum Hall effect. *Phys. Rev. Lett.* **67,** 883–886 (1991).

7. Engel, L. W., Shahar, D., Kurdak, C. & Tsui, D. C. Microwave frequency-dependence of integer quantum Hall-effect — evidence for finite-frequency scaling. *Phys. Rev. Lett.* **71,** 2638–2641 (1993).

8. Hohls, F., Zeitler, U. & Haug, R. Hopping conductivity in the quantum Hall effect: Revival of universal scaling. *Phys. Rev. Lett.* **88,** 1–4 (2002).

9. Hohls, F. *et al.* Dynamical scaling of the quantum Hall plateau transition. *Phys. Rev. Lett.* **89,** 1–4 (2002).

10. Shahar, D. *et al.* A new transport regime in the quantum Hall effect. *Solid State Commun.* **107,** 19–23 (1998).

11. Balaban, N., Meirav, U. & Bar-Joseph, I. Absence of scaling in the integer quantum Hall effect. *Phys. Rev. Lett.* **81,** 4967–4970 (1998).

12. Cobden, D. H., Barnes, C. H. W. & Ford, C. J. B. Fluctuations and evidence for charging in the quantum Hall effect. *Phys. Rev. Lett.* **82,** 4695–4698 (1999).

13. Chklovskii, D. B. & Lee, P. A. Transport properties between quantum Hall plateaus. *Phys. Rev. B* **48,** 18060–18078 (1993).

14. Cooper, N. R. & Chalker, J. T. Coulomb interactions and the integer quantum Hall-effect — screening and transport. *Phys. Rev. B* **48,** 4530–4544 (1993).

15. Ruzin, I., Cooper, N. & Halperin, B. Nonuniversal behavior of finite quantum Hall systems as a result of weak macroscopic inhomogeneities. *Phys. Rev. B* **53,** 1558–1572 (1996).

16. Ilani, S., Yacoby, A., Mahalu, D. & Shtrikman, H. Unexpected behavior of the local compressibility near the B=0 metal-insulator transition. *Phys. Rev. Lett.* **84,** 3133–3136 (2000).

17. Yoo, M. J. *et al.* Scanning single-electron transistor microscopy: Imaging individual charges. *Science* **276,** 579–582 (1997).

18. Yacoby, A., Hess, H. F., Fulton, T. A., Pfeiffer, L. N. & West, K. W. Electrical imaging of the quantum Hall state. *Solid State Commun.* **111,** 1–13 (1999).

19. Zhitenev, N. B. *et al.* Imaging of localized electronic states in the quantum Hall regime. *Nature* **404,** 473–476 (2000).

20. Ilani, S., Yacoby, A., Mahalu, D. & Shtrikman, H. Microscopic structure of the metal-insulator transition in two dimensions. *Science* **292,** 1354–1357 (2001).

21. Efros, A. L. & Ioffe, A. F. Nonlinear screening and the background density of 2DEG states in magnetic field. *Solid State Commun.* **67,** 1019–1022 (1988).

22. Efros, A. L., Pikus, F. G. & Burnett, V. G. Density of states of a two-dimensional electron gas in a long-range random potential. *Phys. Rev. B* **47,** 2233–2243 (1993).

23. Shashkin, A. *et al.* Percolation metal-insulator transitions in the two-dimensional electron system of AlGaAs/GaAs heterostructures. *Phys. Rev. Lett.* **73,** 3141–3144 (1994).

24. Kukushkin, I., Fal'ko, V., Haug, R., von Klitzing, K. & Eberl, K. Magneto-optical evidence of the percolation nature of the metal-insulator transition in the two-dimensional electron system. *Phys. Rev. B* **53,** R13260–R13263 (1996).

**Acknowledgements** We benefited from discussions with A. M. Fink'elstein, Y. Gefen, Y. Meir, A. Stern and N. B. Zhitenev. This work was supported by the Israeli Science Foundation and the German MINERVA foundation.

**Correspondence** and requests for materials should be addressed to S.I. (shahal.ilani@cornell.edu).

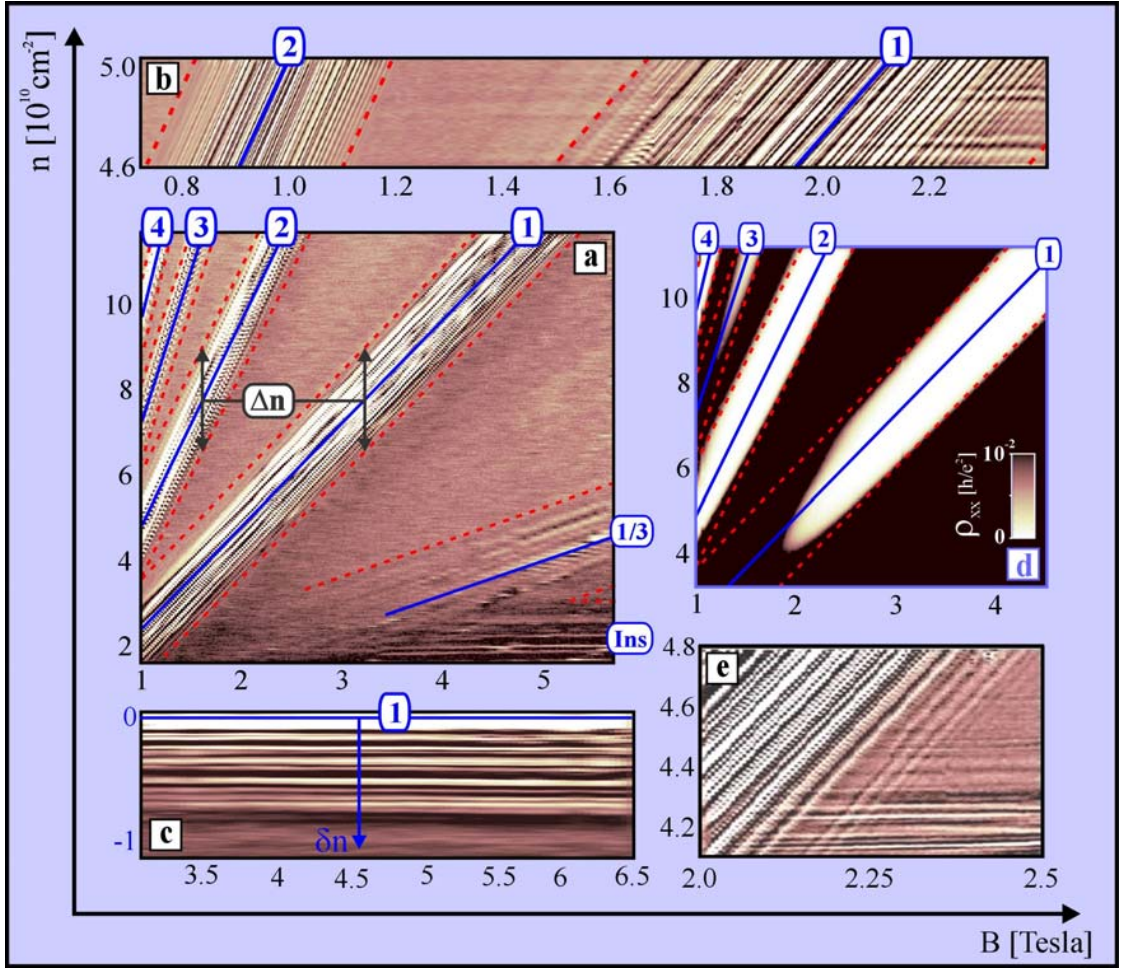

**Figure 1** Measurements of the derivative $d\mu/dV_{BG}$ at an arbitrary position above the 2DES as function of magnetic field ($B$) and density ($n$). This derivative is measured with a single-electron transistor (SET), which offers a spatial resolution of ~1,000 Å and an unprecedented voltage sensitivity of ~1 μV Hz$^{-1/2}$. **a**, A measurement over a large range in $B$ and $n$ demonstrating the alternating pattern of incompressible (bright) and compressible (dark) regions. The bright regions correspond to the QH phases of the system. **b**, Detailed measurement of the $\nu=1$ and $\nu=2$ incompressible regions, revealing a rich fine structure of parallel (black) lines, each representing the charging line of an individual localized state. All charging lines within a certain group have exactly the same slope in the $n$–$B$ plane. **c**, The charging lines around $\nu=1$ plotted as function of $B$ and the density deviation from the full Landau level, $\delta n$. **d**, The longitudinal resistivity, $\rho_{xx}$, measured on the same sample as function of $B$ and $n$. Red lines mark the boundaries, copied from **a**, where localized states disappear. **e**, An overlap region of the $\nu=1$ and $\nu=0$ groups showing that localized states of these different groups coexist and maintain their original slopes.

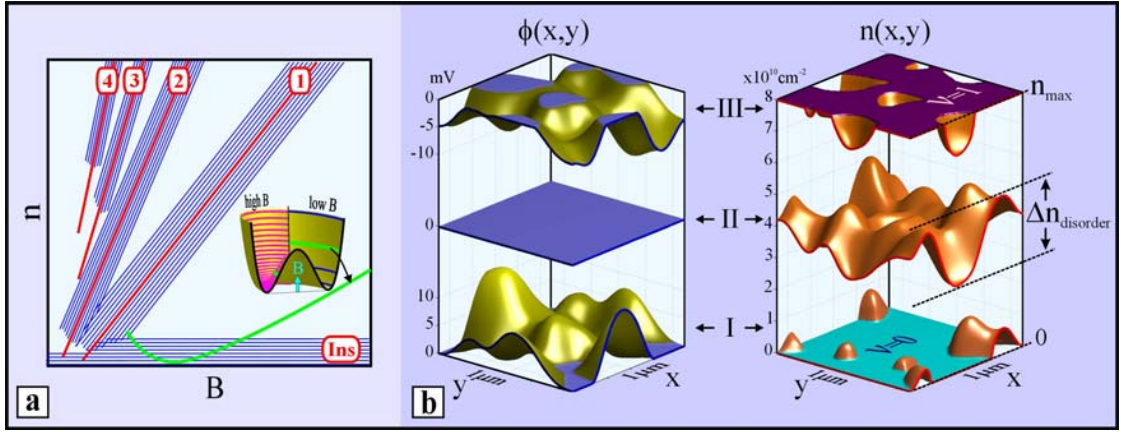

**Figure 2** Theoretical models for localized states. **a**, Schematic illustration of the measured charging lines in the $n$–$B$ plane (blue) compared with a calculated single-particle charging line (green) of a model Mexican hat potential (inset). The single-particle states of the Mexican hat are depicted for low $B$ (blue) and high $B$ (magenta). The specific state that gives rise to the green charging line is marked for both low and high $B$. Its energy approaches the top of the potential hill in the limit of strong $B$. This energy is intermediate between the bottom and top of the disorder potential, and therefore leads to a non-quantized asymptotic slope whose value is between $\nu=0$ and $\nu=1$. The scatter of potential peak heights in a typical disorder will therefore result in charging lines having arbitrary slopes in the range $\nu=0\to1$. **b**, The density profiles, $n(x,y)$, and electrostatic potential profiles, $\phi(x,y)$, calculated self-consistently within the Thomas–Fermi scheme, for an almost-empty (I), half-full (II) and almost-full Landau level (III), in a typical disorder potential. Following ref. 22, the regions in which the electron density is free to fluctuate are taken as infinitely compressible. Near half-filling, the amplitude of the density fluctuations is $\Delta n_{\text{disorder}}$ and the potential is completely screened. Near the bottom of the Landau level ($n=0$) and its top ($n=n_{\text{max}}$) the potential is screened only in compressible pockets, which are surrounded by incompressible regions, marked as $\nu=0$ or $\nu=1$.

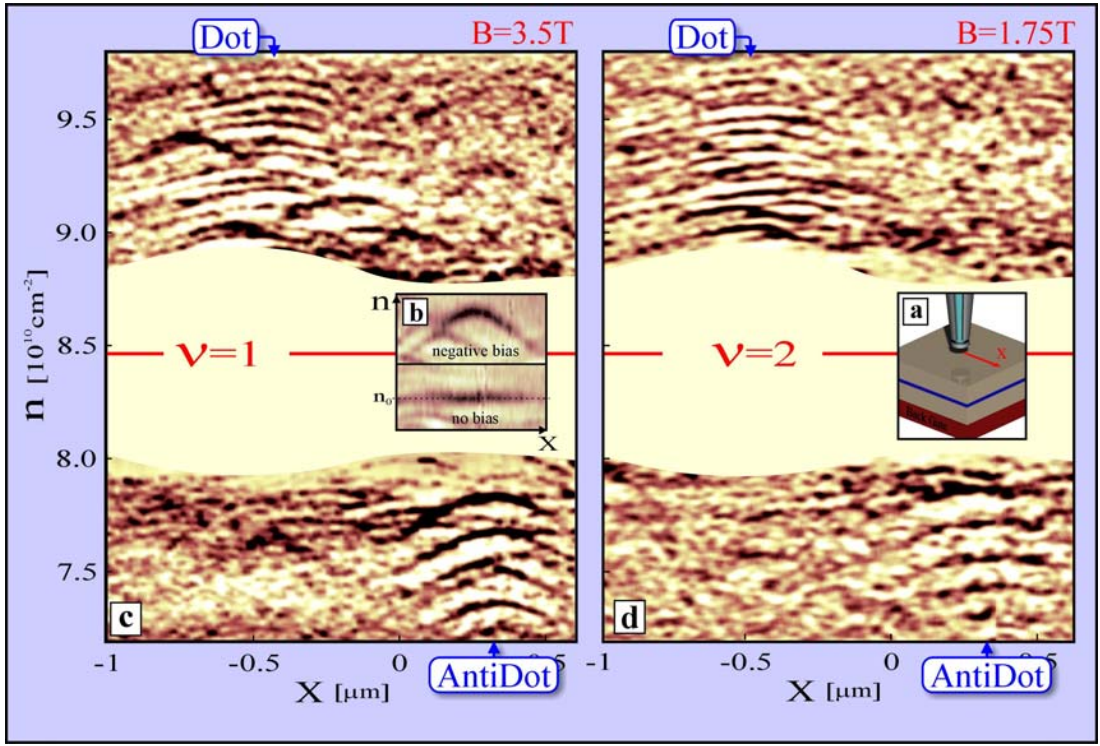

**Figure 3** Spatial scans of localized states. **a**, A SET mounted on a tip of a scanning microscope is used to measure $d\mu/dV_{BG}$ as a function of position, $x$, and density, $n$. **b**, The measured signature of a localized state in the $n$–$x$ plane. The state has a finite spatial extent. Its charging event is thus detected by the SET over a finite range of tip positions. In its non-perturbing state, the tip does not affect the density in which the state is charged ($n_0$). It results in a straight segment parallel to the $x$ axis, which spans the actual spatial spread of the state convoluted with the tip response function. However, if a slight bias is added to the tip, it acts also as a local gate, pushing the charging condition to higher densities in a manner proportional to the local charge density within the localized state. This results in a bent segment whose shape gives additional information on the charge distribution within the individual state. **c, d**, Spectra of localized states measured around $\nu=1$ and $\nu=2$, at the same positions and densities, with a small bias applied to the tip. The localized states bunch spatially into well-ordered families, marked as 'Dot' and 'Antidot'. Despite the difference in the Landau levels and the energy gaps for $\nu=1$ and $\nu=2$, their localized states spectra show remarkable similarity. (The central region in both scans is masked, because, as opposed to the rest of the scan, high resistance of the 2DES prevents the system from equilibrating to the thermodynamic ground state.)

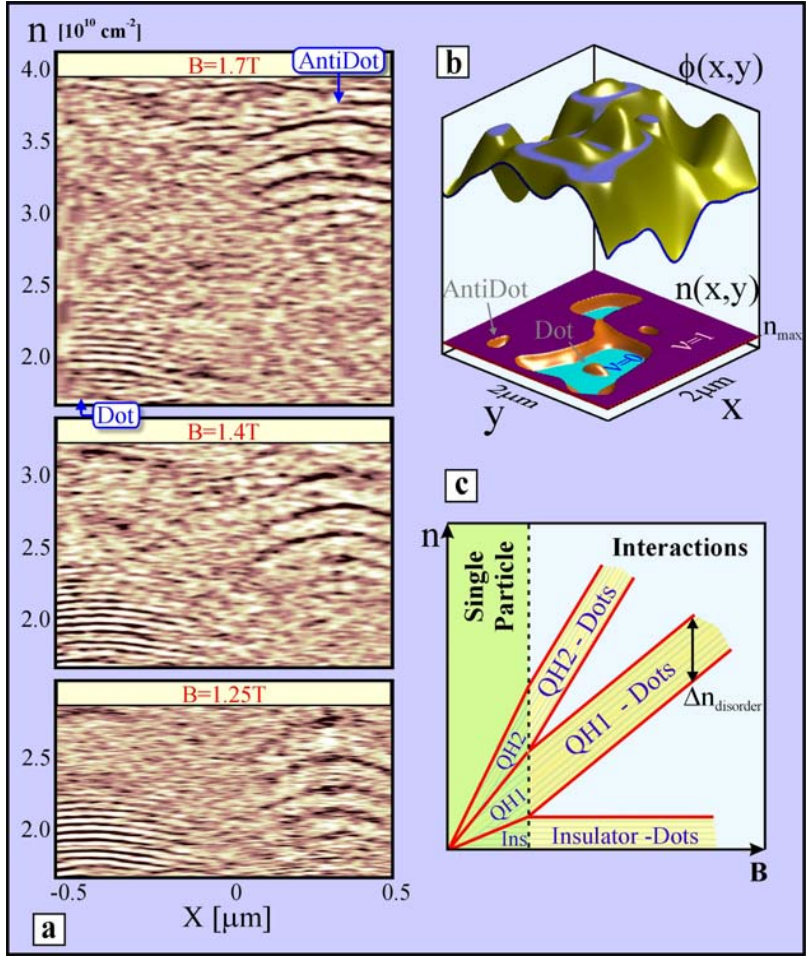

**Figure 4** Localized states in the presence of strong disorder. **a**, Density–position scans at $B$=1.7 T, 1.4 T, 1.25 T tracking a $\nu$=1 antidot and $\nu$=0 dot down to the highly disordered regime at low $B$ ($\Delta n_{disorder} \gg n_{max}$). The highest density in all scans corresponds to $\nu$=1. At the lowest $B$, both dot and antidot coexist—hole-like $\nu$=1 states neighbour electron-like $\nu$=0 states less than 1 μm away. **b**, Self-consistent Thomas–Fermi calculation in the limit of strong disorder. Here, the disorder creates large density fluctuations, which force the Landau level to be either completely empty or completely full and therefore incompressible. Compressible pockets in the $\nu$=0 and $\nu$=1 plateaus result in $\nu$=0 dots and $\nu$=1 antidots, as marked in the figure. On the boundary between the $\nu$=0 and $\nu$=1 regions a narrow compressible strip exists, following a constant energy contour of the bare disorder potential. **c**, A generalized phase diagram for QH-localization as deduced from our measurements. At low $B$, localized states are dominated by disorder and follow the single-particle phenomenology. Here the familiar QH phase transitions at half-filling of Landau levels are observed. At higher fields, interactions prevail and localized states form within dots or antidots. In this regime, new percolation phase boundaries emerge (red lines) and limit the QH phenomena to strips of width $\Delta n_{disorder}$ centred around their corresponding filling factors.